\documentclass{iaus}
\usepackage{graphicx}

\def\aj{\textit{AJ}}
\def\apj{\textit{ApJ}}
\def\apjs{\textit{ApJS}}
\def\mnras{\textit{MNRAS}}
\def\aap{\textit{A\&A}}

\title[The Star Formation History of Late Type Galaxies]
      {The Star Formation History of Late Type Galaxies}

\author[R. Cid Fernandes]{Roberto Cid Fernandes}

\affiliation{Departamento de F\'\i sica, Universidade Federal de Santa
Catarina, Brazil, \break email: cid@astro.ufsc.br}

\pubyear{2004}
\volume{241}
\pagerange{1--8}
\date{?? and in revised form ??}
\jname{Proceedings Title IAU Symposium}
\editors{A.C. Editor, B.D. Editor \& C.E. Editor, eds.}
\begin{document}

\maketitle

\begin{abstract}
The combination of huge databases of galaxy spectra and advances in
evolutionary synthesis models in the past few years has renewed
interest in an old question: How to estimate the star formation
history of a galaxy out of its integrated spectrum? Fresh approaches
to this classical problem are making it possible to extract the best
of both worlds, producing exquisite pixel-by-pixel fits to galaxy
spectra with state-of-the-art stellar population models while at the
same time exploring the fabulous statistics of mega-surveys to derive
the star-formation and chemical enrichment histories of different
types of galaxies with an unprecedented level of detail.  This review
covers some of these recent advances, focusing on results for
late-type, star-forming galaxies, and outlines some of the issues
which will keep us busy in the coming years.  
\end{abstract}

\firstsection 

\section{Introduction}

Understanding the cosmological and internal processes which drive
galaxy evolution is a major goal of contemporary astrophysics.
Empirical information on the star formation history (SFH) of galaxies
is a key piece in this quest.  Impressive progress has been made both
with high $z$ studies, which measure evolution directly by comparing
galaxy properties at different cosmic times, and with SFH recovery
techniques based on color-magnitude diagrams of our closest neighbors,
which recover time information from stellar evolution clocks.  Most of
the data comes from between here and there, where galaxies are neither
far enough to use cosmological clocks nor close enough to resolve
individual stars, and hence SFHs must be retrieved from integrated
light measurements.

Generations of astronomers have worked in this field, and even
limiting the scope of this review to techniques based on optical
spectra and biasing it towards applications to large surveys in the
past few years, so much has been done that it would be impossible to
make justice to all.  This contribution thus presents an inevitably
incomplete review of recent progress in the field. The focus is not so
much on results but mainly on the diversity of methods to go from
integrated optical spectra to SFHs. Browsing through this volume you
will see that so much more is being done as we ``speak'' that I will
close this with a few lines about issues to be explored in the
very near future.

\section{Stellar population mixtures: Ingredients \& observables}

Late type galaxies are evidently composite systems, where multiple
generations of stars contribute to the integrated light. Unlike with
elliptical galaxies, which are often modeled as single age systems,
for spirals and irregulars one cannot evade the challenge of
unscrambling the mixture of photons reflecting different cosmic times,
from the $\sim 10$ Gyr populations of the bulge to the new-born stars
in the disk and starbursting nuclei. This mixture can be represented
by a sum of $N_\star$ populations of different ages and metallicities,

\begin{equation}
\label{eq:synt}
L_\lambda^{gal}(\vec{x},A_V) = 
L_{\lambda_0}^{gal} \sum_{j=1}^{N_\star} x_j l_{\lambda}(t_j,Z_j)
\otimes {\rm LOSVD} \times 10^{-0.4 A_V r_\lambda}
\end{equation}

\noindent where $l_\lambda$ represents the spectrum of population $j$
normalized at $\lambda_0$, $x_j$ is its light fraction and $r_\lambda
= (A_\lambda - A_{\lambda_0}) / A_V$ denotes the reddening law. The
fossile record of the SFH is encoded in the {\it population vector}
$\vec{x}$.  Conversion of this discrete representation to a continuous
one, or from light to mass fractions is straightforward. It is equally
simple to generalize this formalism to allow for population dependent
extinctions, reddening-laws and Line Of Sight Velocity
Distributions. Though this would surely produce a more realistic
model, the task of deriving SFHs from a comparison of
$L_\lambda^{gal}$ with actual galaxy spectra would become so much
harder that this would be an academic refinement at this stage (more
on this in \S\ref{sec:Conclusions}), so lets stick to this simple, yet
useful approach.

Behind its formal simplicity, eq.\ (\ref{eq:synt}) hides a multitude
of astrophysical, mathematical and computational issues which
propagate to a substantial diversity in SFH recovery methods. For
starters, what should one use as the spectral building blocks
$l_\lambda(t_j,Z_j)$?

Bica \& Alloin (1986) and Bica (1988) proposed to work with a base of
observed star cluster spectra, founding a fruitful and inspiring
empirical approach to population synthesis. Since then, the modeling
of Simple Stellar Populations (SSP) has evolved so much that one can
now $\sim$ safely replace observed clusters by theoretical ones with
(arguably) more pros than cons. This major advance came about with the
incorporation of medium--high spectral resolution libraries into
evolutionary synthesis models (Bruzual \& Charlot 2003; Le Borgne
\etal 2004; Gonz\'alez Delgado \etal 2005), which quickly became a
standard ingredient in SFH studies. With the release of new spectral
libraries and evolutionary tracks announced in this conference, fossil
methods now have a long menu of $l_\lambda(t_j,Z_j)$'s to choose
from. Diversity is certainly healthy! Yet, it inevitably brings some
entropy to the field, so it is important to understand how and why
these models differ and how this affects the derived SFHs in
practice. Important steps in this direction have been presented by
Moleva, Panter and Prugniel in this meeting.

Moving away from ingredients towards how to use them, the comparison
of (\ref{eq:synt}) to an observed spectrum can be done either in its
full $\lambda$-by-$\lambda$ power (\S\ref{sec:SpectralSynthesis}) or
in terms of selected {\it spectral indices} such as absorption line
equivalent widths and colors. The latter approach, more common until
very recently, was applied to star-forming galaxies of several kinds
by Raimann \etal (2000); Cid Fernandes \etal (2003); Kong \etal
(2003); Westera \etal (2004). An important result of these studies is
that even the smallest, youngest looking systems contain a mass
dominant $\sim 10$ Gyr population, and thus, contrary to first
impressions, are {\it not} primeval galaxies. (Work based on full
spectral fits confirm this finding and extends it to even more extreme
classes of star-forming galaxies; Corbin \etal 2006; Lisker \etal
2006). Active galaxies of different brands have also been targeted
with such techniques (see Cid Fernandes 2004 for a review).  It is fit
to recall that studies based on a few spectral features often do what
some would call absurd: Fit more populations than observables
available! Those who still have their qualms about algebraic
degeneracy should read Pelat (1998), who, besides clarifying this
issue, proposed an elegant (yet largely overlooked in the literature)
inversion method. Index based work has also entered the new era of
huge databases. Kauffmann \etal (2003) and Gallazzi \etal (2005)
developed a Bayesian technique based on a handful of indices which
does not recover the full time dependent SFH, but provides valuable
estimates on some of its associated ``moments'', like the mean age,
metallicity and fraction of mass formed in recent bursts. Its
application to SDSS data brought important insights on relations
between stellar populations and other galaxy properties, particularly
its mass.

\section{Spectral fits: Methods \& results}

\label{sec:SpectralSynthesis}

\begin{figure}
\includegraphics[width=\textwidth]{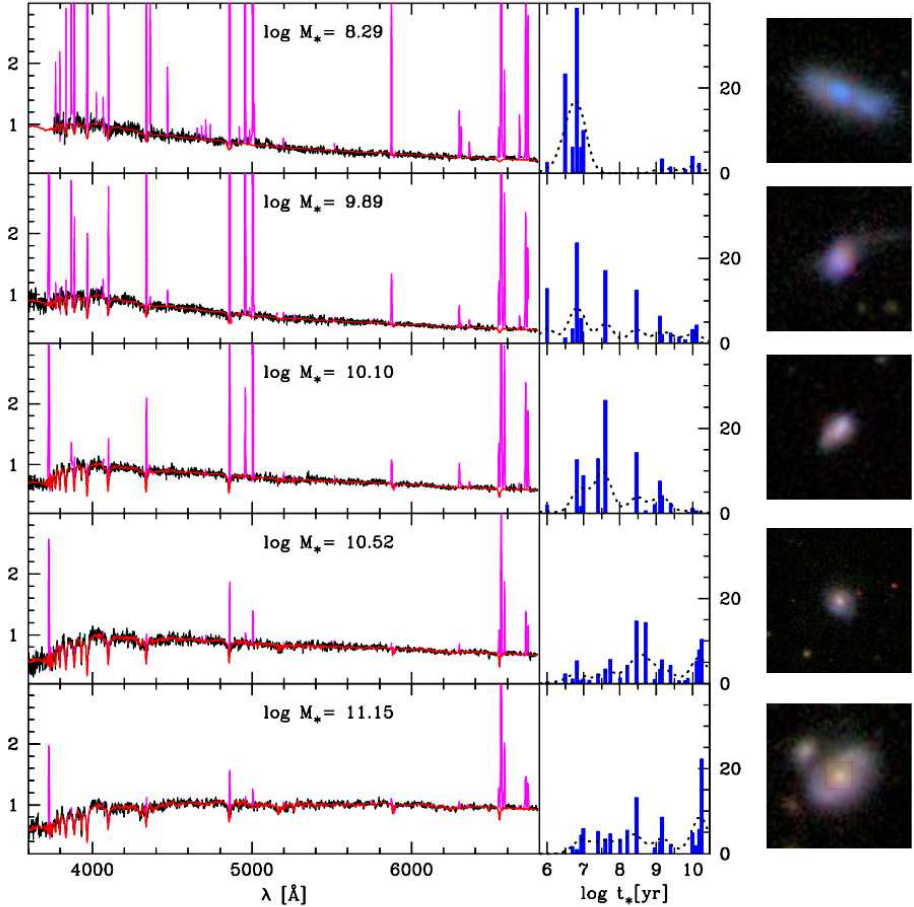}
\caption{\textit{Left:} Observed (black) and fitted (red) spectra of 5
SDSS star-forming galaxies at similar distances (except for the
1$^{st}$, which is closer). Emission line masks are plotted in
magenta. Galaxies are sorted according to their stellar mass,
increasing from $10^{8.29}$ to $10^{11.15}$ M$_\odot$ from top to
bottom. \textit{Middle:} SFH, given in terms of the light fraction at
4020 \AA\ associated with each of the 25 ages included in the fits and
marginalizing over $Z$. Dotted curves show a 0.5 dex gaussian-smoothed
version of the population vector $\vec{x}$.  Notice how the balance
between recent and past star-formation changes in pace with the
$M_\star$ sequence. \textit{Right:} SDSS image.}
\label{fig:fits}
\end{figure}

The new vintage of high spectral resolution evolutionary synthesis
models offered the long awaited possibility of performing detailed
``\AA-by-\AA'' fits of galaxy spectra. Fig.\ref{fig:fits} illustrates
that this is no longer just a possibility, but a reality.

Emission line studies have welcome this advance.  Before it, the
traditional approach to clean up starlight for emission line
measurements was to subtract a suitable template galaxy, but as soon
as your target contains a young population (as is the case of late
types) it becomes impossible to find a suitable template which does
not have emission lines of its own. Let me open a parenthesis to
illustrate that this is not as minor a problem as it may seem to this
audience. For $\sim$ 2 decades studies of Seyfert 2 nuclei used
elliptical galaxy templates to represent the stellar spectrum, which
often produced a seemingly featureless residual continuum (see Sarzi's
contribution for an update on this). This was first attributed to
accretion disk light, then to scattered photons from a hidden nucleus,
but these interpretations clashed head-on with other pieces of AGN
phenomenology. It took many papers and telescope time to realize that
these nuclei are often surrounded by stars much younger than those
found in ellipticals, and that the mysterious residual continuum was
essentially a side effect of a template-mismatch. All this work
(including mine!) would have been superfluous if the ingredients for
spectral fits like those used in Fig.\ \ref{fig:fits} existed back
then. The main message here is that progresses in stellar population
modelling have an impact well outside the stellar population field.

Our interest here, of course, is not to get rid of stellar photons,
but to retrieve the SFH information they carry in a peculiarly
scrambled way. The fits in Fig.\ \ref{fig:fits} were constructed
combining SSPs of 25 different $t_j$'s and 6 $Z_j$'s from the BC03
models, all extincted by the same screen of Galactic-like dust, and
adjusting the velocity dispersion $\sigma_\star$ as well as possible
velocity off-sets with respect to the rest-frame. Emission lines and
bad pixels are masked from the fits, which minimize a standard
$\chi^2$ figure of merit.

Fitting all pixels saves you the trouble of picking and measuring
indices. On the other hand, the factor of $\sim 1000$ increase in the
number of observables slows computations, though this can be handled
efficiently with mathematical and programming tricks. Apart from such
minor technical differences, the fundamental issues faced by
index-based and full spectral synthesis methods to recover SFHs are
the same, like: which $l_\lambda(t_j,Z_j)$ building blocks to use, how
to go from the observables to the parameters, how to handle
astrophysical and mathematical degeneracies and uncertainties, and to
which degree can one trust the resulting SFHs. Let us browse through
some of these inter-related points.

It is intuitively obvious that the $N_\star = 150$ populations used in
Fig.\ \ref{fig:fits} cannot be trusted individually. This
over-dimensioned parameter space, philosophically rooted in a
``principle of maximal ignorance'', must somehow be {\it compressed}
to produce SFHs with only as much resolution as afforded by the data.
This issue has a long history in index methods. In the context of
spectral fits, it was first seriously tackled by the MOPED group, who
developed a data compression method which preserves information on the
SFH (Heavens \etal 2000; Panter \etal 2003; Reichhardt \etal 2001;
Mathis \etal 2006). The STECMAP group (Oczvirk \etal 2006) analyzed
this issue from a different perspective, and proposed a regularization
technique which controls the smoothness of the resulting SFHs in a
data driven fashion, while STARLIGHT (the code used to produce Fig.\
\ref{fig:fits}; Cid Fernandes \etal 2005) works with an oversampled
population vector $\vec{x}$ all the way through the fit, and only then
rebins or smooths spectrally similar components onto a coarser but
more robust $\vec{x}$.  At the risk of oversimplifying the issue,
these 3 examples could be described as {\it a priori}, ``on the fly''
and {\it a posteriori} compression approaches,
respectively. Experiments indicate that the age resolution achievable
with optical spectra of realistic S/N is somewhere between $\Delta
\log t = 0.5$ and 1 dex, which even in the worst case represents a
great improvement over a SFH description based only on mean ages.

MOPED, STARLIGHT and STECMAP are just examples of the booming business
of spectral synthesis codes. A staggering variety of techniques are
being explored, including active instance-based machine learning
(Solorio \etal 2005), convex algebra (Moultaka 2005), PCA (Li \etal
05), Bayesian latent variable modelling (Nolan \etal 2006), direct
fitting (Tadhunter \etal 2005; Moustakas \& Kennicutt 2006; Walcher
\etal 2006), and others (see also MacArthur, van der Marel and Sarzi
contributions.) This diversity may look scary, so let me remind the
reader that all these methods share a same goal, namely, to map the
space of observables to the SFH parameters, and so should produce
similar results despite differences in formalism, elegance, complexity
and speed. Though more tests are desirable, the amazing agreement
between MOPED, STARLIGHT and STECMAP results for the challenge
proposed by the organizers of this meeting suggests that algorithm
should {\it not} be considered another free parameter in fossil
methods.

This convergence of independent codes is even more reassuring when we
consider that, besides algorithm, they also differ in dozens of more
subtle, yet relevant details, some more technical, others more
astrophysical. ``Technical'' differences include data pre-processing
steps, handling of kinematical parameters, choice of extinction-law
and whether $A_V < 0$ is allowed or not (see Gallazzi \etal 2005 and
Mateus \etal 2006 for more on this curious point), whether pure SSPs
or constant star-formation episodes within time bins are used for
$l_\lambda(t_j,Z_j)$, which time-bins are used and whether
cosmological consistency ($t < 14$ Gyr) is enforced {\it a priori},
whether the continuum is fitted or rectified, emission line masks,
non-stellar components such as nebular emission and AGN, etc.

A more astrophysically relevant difference is the handling of
metallicities. Some algorithms work at fixed $Z$, some impose simple
$Z$-$t$ relations, some allow one $Z$ per $t$-bin in a non-parametric
way, and others treat $Z$ and $t$ independently. The information on
{\it chemical evolution} retrievable by the synthesis clearly depends
on this choice. As a matter of fact, unlike the stellar mass assembly
histories retrieved by fossil methods, which are getting a lot of
(well deserved) attention, not much has been done on the way of
chemical evolution, probably because of fears that degeneracies plus
noise would kill the $Z(t)$ signal. On the other hand, Cid Fernandes
\etal (2005) and Gallazzi \etal (2005) showed that fossil methods
applied to SDSS data do recover mean stellar metallicities which
behave in an astrophysically expected manner when correlated with
stellar mass and gas-phase metallicity, showing that, in agreement
with test-simulations, at least the first moment of the $Z$
distribution is well recovered. This lead us to venture into the next
step, i.e., checking whether fossil methods recover reasonable
chemical evolution patterns.

\begin{figure}
\includegraphics[height=12cm,width=\textwidth]{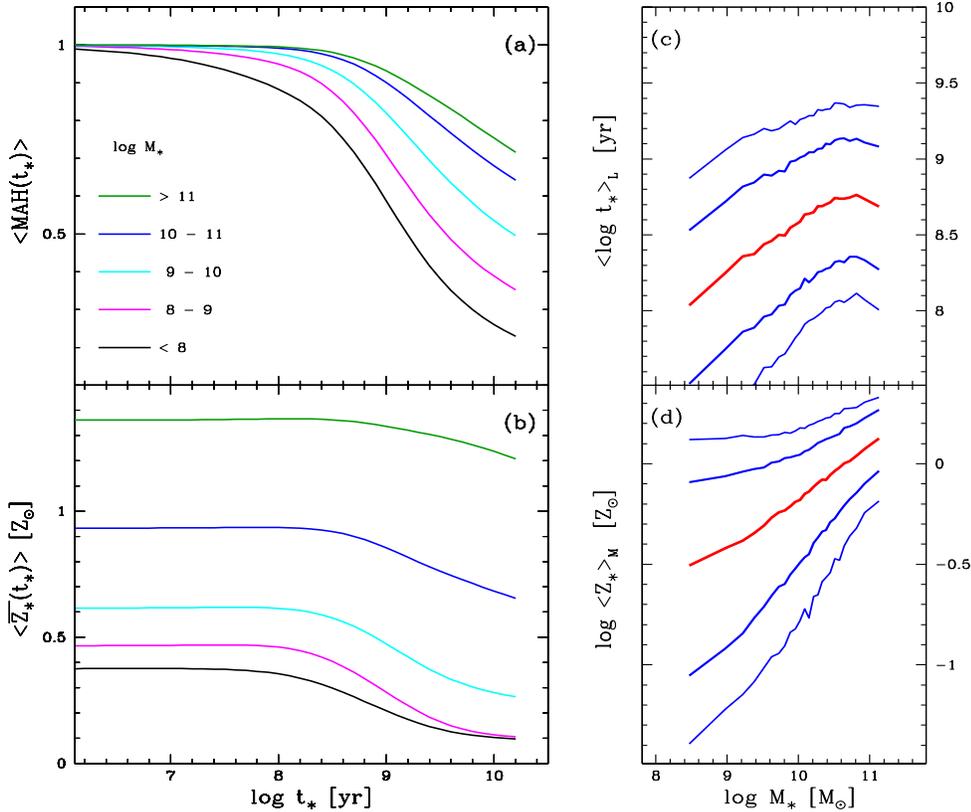}
\caption{Normalized Mass Assembly Histories \textit{(a)} and evolution
of the mean stellar metallicity \textit{(b)} for 84828 star-forming
galaxies from the SDSS, averaged over 5 mass intervals. Panels
\textit{(c)} and \textit{(d)} show the relation between light and
mass-weighted first moments of the $\log t$ and $Z$ distributions and
the stellar mass.  The lines show the median, 1 and 2 sigma equivalent
percentiles in 25 equally populated $M_\star$-bins. (Adapted from
Asari \etal 2007, in prep.).}
\label{fig:chem_evol}
\end{figure}

Fig.\ \ref{fig:chem_evol} illustrates results for a sample of
star-forming galaxies.  The bottom-left panel shows the {\it time
dependent} mean stellar metallicity averaged over galaxies in 5 mass
ranges. Besides the fact that lower $M_\star$ galaxies have lower $Z$
nowadays, the curves show that chemical evolution proceeds at a slower
pace the smaller $M_\star$ is, with the lower $M_\star$ bin reaching
its present day stellar metallicity as recently as $\sim 10^8$ yr ago,
whereas in massive galaxies all evolution occurs early and within
$\sim 1$ age-resolution element. The mass assembly histories retrieved
from the spectral fits show this same speed-mass downsizing
pattern. The panels on the right show the less detailed (but still
useful) first moments of the $t$ and $Z$ distributions, deliberately
mixing light and mass-weighted quantities to remind the reader of the
various ways in which the synthesis results can be manipulated in the
interpretation stage. An important ``detail'' is that the growth
curves in Fig.\ \ref{fig:chem_evol} were smoothed by $\Delta \log t =
1$ dex, which yields robust results while still providing a reasonably
detailed picture of evolution. No such compression was applied in Cid
Fernandes \etal (2006), where these same curves were smoothed by just
enough to disguise the discreteness of our base while not hiding
features like humps close to ages of 1 Gyr. Such fine details reflect
deficiencies in the base (e.g., mismatch between spectroscopic and
stellar evolution $Z$'s) or artifacts of the method, and, while not
changing the general picture, serve to remind us of our limitations
and fight temptations of over-interpreting the fits.  Notwithstanding
these caveats, this exercise demonstrates that fossil methods are
mature enough to contribute to chemical evolution studies.

\section{Summary and outlook}

\label{sec:Conclusions}
This highly incomplete overview tried to outline recent progresses in
studies which recover the SFH of galaxies out of integrated optical
spectra. Luckily, other contributions in this volume expand upon points
compressed beyond recognition in these pages.  20 years ago Searle
(1986) expressed the reigning skepticism towards this topic,
classifying it as ``\ldots a subject with bad reputation. Too much has
been claimed, and too few have been persuaded.'', an opinion shared by
many (including myself) up to not long ago. Since then, ingredients
and methods have evolved to a point that it became impossible to deny
the power of synthesis techniques as a tool to bridge the gap between
the fabulous data sets available nowadays and the ever more
sophisticated stellar population models.  A variety of techniques have
been applied in the reconstruction of the mass assembly and even the
chemical enrichment histories of galaxies, leading to important
constraints for galaxy evolution scenarios.

These optimistic words should not convey the idea that all is done!
After all, despite the ``long and venerable history'' (Searle's words
again) of applied population synthesis, fossil methods of the kind
discussed in this review have practically re-started from scratch in
the past $\sim 3$ years. One should thus remain cautious and skeptic
until this field reaches full maturity. It is hard to say when this
will happen, but some of the hurdles on the way are clear, so lets
indulge in a short futurology exercise.

First, as we have seen, there is a strong drive towards full spectral
fits. These will gather even more momentum with the imminent release
of new $\sim$ \AA-resolution evolutionary synthesis models.  One does
not need a crystal ball to foresee that the combination of new
libraries, the proliferation of synthesis methods and the already
abundant data will fill up hundreds of journal pages in the new couple
of years, reporting results whose compatibility will not be trivial to
assess at first. The introduction of $\alpha$-enhanced libraries
(discussed in Coelho's talk), in particular, will add a new and
qualitatively different dimension to parameter space, inevitably
increasing complexity. Like a spoilt kid with too many toys to chose,
this massive overdosis may throw us into a temporary state of
confusion, which is why comparative studies (in the spirit of the
challenge posed to participants of this meeting) would be highly
desirable.

Secondly, the agreement between different methods suggests that we are
reaching the limit of information that can be extracted from optical
spectroscopy. Indeed, some of the methods are designed to do exactly
this! It thus seems unlikely that different methodologies will bring
substantial improvements to the $t$ and $Z$ resolutions of currently
existing SFH recovery tools. After fitting every single optical pixel,
it is clear that progress will require stretching the spectral horizon
(say, to the near-IR range explored in Lan\c{c}on's talk). This brings
in new challenges and difficulties, specially for those of us who have
grown accustomed to the comparatively easy life of optical astronomy.

Finally, a few words on what seems to be the major fly in the ointment
of current fossil methods: Dust. Equation (\ref{eq:synt}), used in one
way or another by most spectral synthesis codes, pictures a galaxy as
a clean system of stars seen behind a sheet of dust.  Real galaxies
are not quite like that, particularly non ellipticals. At the very
least, one should allow the younger populations to be dustier than the
others (Charlot\& Fall 2000). This has been done in index-methods
which compare data to a large library of precomputed models (eg,
Kauffmann \etal 2003), as well as in applications where the number of
populations is relatively small (eg., Poggianti \etal 2001; Mayya
\etal 2004; Solorio \etal 2005), such that the parameter space can be
held under control. However, attempts to retrieve more than one value
of $A_V$ from general full spectral synthesis in the optical as the
ones described in \S\ref{sec:SpectralSynthesis} have stumbled upon
less than satisfactory results, with simulations indicating that the
extra extinction is not well recovered. This is not surprising given
the nasty non-linearities and degeneracies which come together with
more complete modeling. Interestingly, single $A_V$ fits of
star-forming galaxies in the SDSS find that the line emitting regions
are $\sim$ twice as extincted as the stellar population as a whole
(Cid Fernandes \etal 2005), in excellent agreement with detailed
studies of nearby galaxies (Calzetti \etal 2004). This is rather
ironic, though, since HII regions are also where the youngest,
ionizing populations reside, and thus the result that $A_{V,gas} \sim
2 A_{V_\star}$ indicates that the fits should have allowed at least
part of the $t < 10$ Myr stars to suffer twice as much extinction as
the others! Warnings about such problems have been issued long ago
(Witt \etal 1992), but optical ``synthesizers'' still have not come up
with a fully satisfactory way to deal with them.  It is unclear to
which extent naive modelling of dust effects is affecting SFH studies
of late types in general, but it is clear that the dustier beasts
(like LIRGS and ULIRGS) definitely need a more refined treatment. It
is also unclear how much improvement can be made with optical data
alone. Going back to my previous point (and judging from Bressan's
contribution and the GRASIL group work; e.g., Silva \etal 1998), going
beyond optical is inevitable.  Combining the kind of detailed optical
synthesis discussed here with wider-scale SED modeling involves much
more than simply adding more $\lambda$'s to the fits, and so should
keep us busy for a long time.

\begin{acknowledgments}
I thank the organizers of this meeting for the invitation, the SDSS
team and members of the Semi Empirical Analysis of Galaxies (SEAGal)
collaboration: N. Asari, J. M. Gomes, A. Mateus, J. P. Torres-Papaqui,
W. Schoenell, L. Sodr{\'e}, G. Stasi{\'n}ska.
\end{acknowledgments}

\end{document}